\documentclass[prb,twocolumn,floatfix,longbibliography,superscriptaddress]{revtex4-2}
\usepackage[english]{babel}
\usepackage[utf8]{inputenc}
\usepackage[T1]{fontenc}
\usepackage{amsmath}
\usepackage{amssymb}
\usepackage{graphicx}
\usepackage{bm}
\usepackage{color}
\usepackage{hyperref}
\usepackage{booktabs}
\usepackage{upgreek}
\usepackage{scalerel}
\usepackage{pgffor}
\usepackage{pdfpages}
\usepackage{float}
\usepackage{appendix}
\usepackage{physics} 
\usepackage{lscape}   
\usepackage{silence}
\usepackage{tikz}
\usepackage{soul,xcolor} 
\usepackage[normalem]{ulem}
\setstcolor{red} 

\usepackage{silence}
\usepackage{comment}  
\makeatletter
\AtBeginDocument{\let\LS@rot\@undefined}
\makeatother
\begin{document}
\title{Identifying trivial and Majorana zero-energy modes using the Majorana polarization}
\author{Oladunjoye A. Awoga}
\email[  ]{oladunjoye.awoga@ftf.lth.se}
\affiliation{Division of Solid State Physics and NanoLund, Lund University, Box118, 22100 Lund, Sweden}
 \author{Jorge Cayao} 
 \email[ ]{ jorge.cayao@physics.uu.se}
\affiliation{Department of Physics and Astronomy, Uppsala University, Box 516, S-751 20 Uppsala, Sweden}

\date{\today}
\begin{abstract}
 In this work we consider superconductor-semiconductor hybrids containing both trivial and Majorana zero modes and explore their signatures in the  Majorana polarization. In particular, we consider trivial zero energy states due to confinement and disorder, which seem to be very likely experimental scenarios.  We show that the Majorana polarization is able to characterize the topological phase transition as well as the emergence of Majorana zero modes even when trivial zero-energy states proliferate.  Notably, the Majorana polarization inherits direct information about spatial correlations which are then the key for distinguishing Majorana and trivial zero-modes. We demonstrate the utility of the Majorana polarization in normal-superconductor junctions and superconductor-normal-superconductor Josephson junctions. Our results support the interpretation of the Majorana polarization as a real space topological indicator.
\end{abstract}

\maketitle

\section{Introduction}
\label{section:Intro}
Majorana zero modes (MZMs)  emerge as zero-energy edge states in  topological superconductors, a topological phase that has been predicted to occur in superconductor-semiconductor systems at strong Zeeman fields \cite{Leijnse_2012,sato2016majorana,sato2017topological,Aguadoreview17}.  Since its conception, this idea has motivated an impressive amount of  theoretical and experimental works searching for Majorana signatures \cite{Tanaka2012Symmetry,frolov2019quest,flensberg2021engineered,Barman2021Symmetry,Marra_2022,tanaka2024theory} but it is still unclear whether MZMs have been observed or not \cite{prada2019andreev}.   The main issue was shown to be the proliferation of  trivial zero-energy states (TZESs) well below the expected topological phase transition \cite{Bagrets:PRL12,Pikulin2012A,PhysRevB.86.100503,PhysRevB.91.024514,PhysRevB.86.180503,PhysRevB.98.245407,DasSarma2021Disorder,PhysRevB.104.134507,PhysRevB.105.144509,PhysRevB.107.184509,baldo2023zero,PhysRevB.107.184519,Pal2024Honing}, with signatures  similar to those due to MZMs \cite{TK95,KT96,Kashiwaya_RPP,PhysRevLett.98.237002,PhysRevLett.103.237001,PhysRevB.82.180516}. This in turn has motivated intense efforts aiming at distinguishing between MZMs and TZESs \cite{PhysRevB.97.161401,PhysRevLett.123.117001,PhysRevB.104.L020501,baldo2023zero,aghaee2022inas,vimal2024entanglement} but the problem is far from being settle \cite{prada2019andreev}. 

\begin{figure}[!t]
	\centering
	\includegraphics[width=0.4\textwidth]{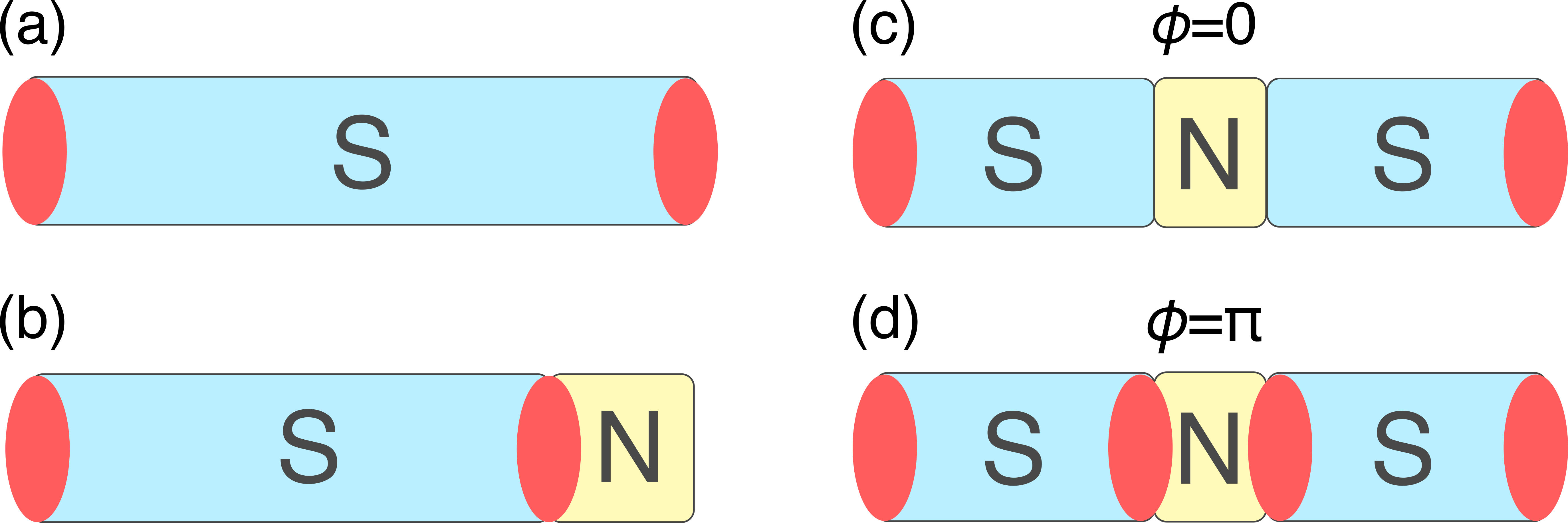}
	\caption{Sketch of the three superconducting systems with SOC and  Zeeman field considered in this work. Panel (a) shows a superconductor S, while (b) a SN  junction, in both cases with MZMs at the edges of S. Panels (c,d) depict a short SNS junction with two MZMs at $\phi=0$ and four MZMs at $\phi=\pi$. The superconducting region (S) is indicated by cyan color, while the normal region (N) by yellow color.}
	\label{Fig1} 
\end{figure}

In this work we show that the Majorana polarization, which characterizes the expectation value of the particle-hole operator, can be used as a real space topological indicator that   distinguishes between MZMs and TZESs in superconducting systems [Fig.~\ref{Fig1}]. The Majorana polarization was used before to characterize MZMs but in the absence of TZESs  \cite{sticlet.bena.12,sedlmayr.bena2015,Sedlmayr2016Majorana,Bena2017,Kaladzhyan2017majorana}. The Majorana polarization was initially defined as an absolute value function and involved only over one half of the  system. As such, this initial definition of the Majorana polarization is unable to distinguish between MZMs and TZESs and cannot capture the nonlocal correlations between MZMs (Majorana nonlocality) which involves having both ends of the system. We extend the previous definition and introduce the Majorana polarization as quantity that captures the correlation between MZMs emerging at the edges, and eliminate the use of absolute value. To show the usefulness of the Majorana polarization, we first  consider a finite superconductor with spin-orbit coupling  under the presence of a Zeeman field and demonstrate that it detects the topological phase transition and the emergence of   MZMs. In this case, we find  that the Majorana polarization exactly captures the behavior of the Pfaffian, which then enables us to interpret the Majorana polarization  as a real space topological indicator. Notably, we discover that the Majorana polarization remains a good topological invariant even under strong scalar disorder, where the visibility of the gap closing is degraded and an avalanche of disorder-induced TZESs emerge. 

We then confirm the utility of the Majorana polarization to distinguish between  MZMs and TZESs induced by confinement in  normal-superconductor (NS) junctions and superconductor-normal-superconductor  (SNS) Josephson junctions. In the SNS junction we also find that the Majorana polarization reveals the emergence of the two MZMs at $\phi=0$ and four MZMs at $\phi=\pi$, which further support the interpretation of the Majorana polarization as a real space topological invariant. Since disorder and  confinement effects represent two mechanisms that induce TZESs, and very likely   present under realistic conditions, our results can be helpful for predicting the topological superconducting phase and MZMs  in  superconductor-semiconductor hybrid systems.

The remainder of this work is organized as follows. In Section~\ref{sec2} we discuss the model and method,  while in  Section~\ref{sec3} we obtain the Majorana polarization in a finite nanowire without and with disorder. In Sections \ref{sec4} and \ref{sec5} we show how the Majorana polarization distinguishes between TZESs and MZMs in NS and SNS junctions, respectively.  Finally in Section~\ref{sec6}, we present our conclusions. For completeness, in Appendix \ref{App:MinimalKitaev} we address the Majorana polarization in a minimal Kitaev chain, while in Appendix \ref{app:MajPol_SingleDis} we compare    the Pfaffian and Majorana polarization in a disordered system.


\section{Models and method}
\label{sec2}
We consider three one-dimensional (1D) superconducting systems, including a finite superconductor, a NS junction, and a short SNS Josephson junction, as schematically shown in Fig.\,\ref{Fig1}.  We model these system by the following tight-binding Hamiltonian 
\begin{equation}
\label{Hmodel}
	\begin{split}
\mathcal{H}_{}=&\sum_{j=1, \sigma \sigma^{\prime}}^{M} d_{j \sigma}^{\dagger} \left[\varepsilon_{\mathrm{}}(j)\delta_{\sigma \sigma^{\prime}} + B \sigma_{\sigma \sigma^{\prime}}^x+i\Delta(j) \sigma_{\sigma \sigma^{\prime}}^y\right] d_{j \sigma^\prime}\\
&+\sum_{j=1, \sigma \sigma^\prime}^{M-1}  d_{j \sigma}^{\dagger}\left[-t_{\mathrm{}}\delta_{\sigma \sigma^{\prime}} +\alpha_{\mathrm{}}\sigma_{\sigma \sigma^{\prime}}^y\right]d_{j+1, \sigma^\prime}+\text { H.c. },\\
\end{split}
\end{equation}
where  $\varepsilon(j)=2 t-\mu(j)+V(j)$ is the site-dependent onsite energy with $t$ being the   nearest neighbor hopping amplitude, while $\mu(j)$  is the spatial dependent chemical potential and   $V(j)$ a spatial dependent potential characterizing  scalar disorder, which will be specified later. Here $d_{j\sigma}$ and $d^{\dagger}_{j\sigma}$, respectively, destroy and create a fermionic state with spin $\sigma$ at site $j$ in the superconducting system and runs over the entire system of length $L=Ma$, being $M$ the total number of sites and $a$ the lattice spacing.  Moreover, in Eq.\,(\ref{Hmodel})   $\alpha_{\mathrm{}}$ is the SOC strength, $B$ the Zeeman field along $x$, and $\Delta(j)$ the space dependent superconducting pair potential.  Thus, the finite superconductor  in Fig.\,\ref{Fig1}(a) of length $L$ is described by an homogeneous pair potential $\Delta(j)=\Delta$ and finite chemical potential $\mu(j)=\mu$; the potential $V(j)$, however, still depends on space, see below. In this single superconducting nanowire, a topological phase transition occurs at  $B=B_{\rm c}$ where $B_{\rm c}=\sqrt{\Delta^{2}+\mu^{2}}$. Then, MZMs emerge when  $B>B_{\rm c}$, where the system becomes a $p$-wave superconductor and effectively describes the physics of the Kitaev chain~\cite{kitaev2001unpaired}.

To model the  NS and SNS junctions in Fig.\,\ref{Fig1}(b,c),   the N region  is considered to have $\Delta=0$ and $\mu_{\rm N}$, while the S region $\Delta\neq0$ and $\mu_{\rm S}$. We note that in the case of the SNS Josephson junction, we also consider a finite phase difference between the pair potentials of the S regions, such that we can explore the Josephson effect. The lengths of N (S) for these junctions are denoted as $L_{\rm N(S)}=M_{\rm N(S)}a$, where $M_{\rm N(S)}$ correspond to the number of sites in the N(S) regions. Then, the S regions can become topological with MZMs at their ends for $B>B_{\rm c}$

We are interested in exploring the Majorana polarization to distinguish between  MZMs and TZESs. Before discussing how to obtain the Majorana polarization, we point out that  to characterize the topological phase of the Rashba model given by Eq.\,(\ref{Hmodel}), it is common to use the Pfaffian topological invariant given by \cite{nadj2013proposal,cheng2021fatemajoranazeromodes}
\begin{equation}
\label{eq:Pf}
	\mathcal{M} = {\rm sgn}[{\rm Pf }[\tilde{\mathcal{H}}]]
\end{equation}
where {\rm sgn} represents the sign operation, {\rm Pf} denotes the Pfaffian, and $\tilde{\mathcal{H}}$  is a skew-symmetrix matrix obtained by writing the Hamiltonian in Eq.~\eqref{Hmodel} in the  Majorana basis, ~\cite{kitaev2001unpaired,nadj2013proposal,Liu2021Fate,awoga2024controlling}. We note that $\mathcal{M}$ is sometimes referred to as Majorana number. To set a starting point in this work, we will use the Pfaffian to characterize the topological phase  of the superconductor with homogeneous pair potential and modeled by Eq.\,\eqref{Hmodel}. For this purpose, $\mathcal{M}$ will  be obtained in real space by taking periodic boundary conditions (closed system) such that the nanowire becomes a ring~\cite{awoga2024controlling}. Thus, when MZMs emerge at the edges, under periodic boundary conditions, they hybridize and acquire finite energy.  Furthermore, it is worth noting, however, the Pfaffian captures the fermion parity of the ground state. Thus, every time an energy level crosses zero, the parity of the system changes and therefore the  Pfaffian changes sign, which is expected irrespective of either the trivial or topological nature of the system. Thus, in spite of the benefits of the Pfaffian as a bulk topological invariant, there is a need for an alternative metric to characterize the topological properties in real space, specially when the system is inhomogeneous and TZESs are present. 

We argue that the issue with the Pfaffian in inhomogenous systems can be resolved by using   the Majorana polarization, which, as we show in this work, turns out to be a more powerful topological indicator because it allows to distinguish between  MZMs and TZESs. For this reason, we start by defining  the Majorana polarization at each site $j$ as \cite{sedlmayr.bena2015}
\begin{equation}
\label{eq:MP}
 P_{j} \equiv \frac{\langle \Psi_j| \mathcal C| \Psi_j \rangle}{\langle \Psi_j | \Psi_j \rangle} = \frac{\sum_{\sigma=\uparrow,\downarrow}2 u_{j\sigma}^{(0)}\big[v_{j\sigma}^{(0)}\big]}{\sum_{\sigma=\uparrow,\downarrow}\big|u_{j\sigma}^{(0)}\big|^2+\big|v_{j\sigma}^{(0)}\big|^2}\,,
\end{equation}
where $\mathcal C = \tau^x\otimes\sigma^0 \mathcal{K}$ is the particle-hole operator, with $\mathcal{K}$ the conjugation operator, $\tau^{i}$ and $\sigma^{i}$ the $i$-th Pauli matrix in Nambu and spin spaces, respectively, while $u_{j\sigma}^{(0)}$ and $v_{j\sigma}^{(0)}$ represent the electron and hole components  with spin $\sigma$ of the wavefunction $\Psi_j$ at site $j$ of the lowest energy state. At this point, we anticipate  that in a finite topological superconductor, the Majorana polarization $P_{j}$ of the left and right halves of the system develop opposite signs. We show this effect by taking a two-site Kitaev chain at the sweet spot where MZMs appear, see Appendix \ref{App:MinimalKitaev}. Motivated by this fact, we take  Eq.\,(\ref{eq:MP}) and define the Majorana polarizations associated to the left and right half of the system, which we refer to as left (L) and right (R) Majorana polarizations and obtain them as,
\begin{equation}
\label{MPMP}
\begin{split}
P_{\rm L}&=\sum_{j=1}^{j=M/2} P_{j}\,,\\
P_{\rm R}&=\sum_{j=M/2+1}^{j=M} P_{j}\,.
\end{split}
\end{equation}

  \begin{figure*}[!t]
\centering
	\includegraphics[width=1\textwidth]{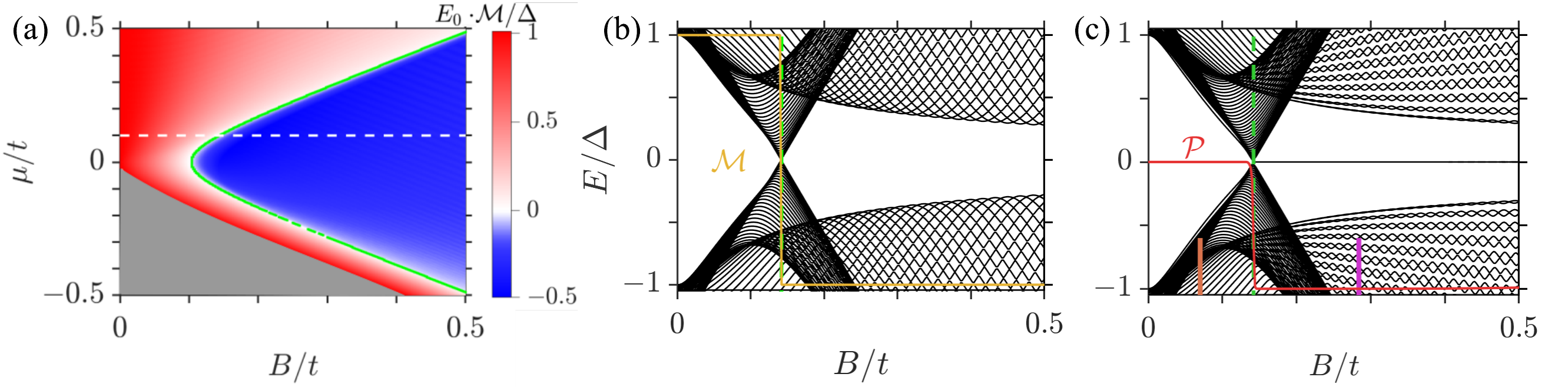}
\caption{Topological phase and low-energy spectrum for a superconducting nanowire. (a) Pfaffian $\mathcal{M}$ from Eq.\,(\ref{eq:Pf}) obtained for a closed system as a function of the chemical potential $\mu$ and Zeeman field $B$, with the intensity given by $E_0\cdot \mathcal{M}/\Delta$, where $E_{0}$ corresponds to the lowest energy level. Here, the blue and red regions depict the nontrivial and trivial topological phases, respectively,  while the white region indicates the phase transition.  The grey region corresponds to $E_{0}/\Delta>1$.  The green curve in (a) corresponds to  the topological phase transition line obtained from the Majorana polarization $\mathcal{P}$ in an open system,  showing great agreement with the Pfaffian predicted phase transition.   (b) Low-energy spectrum  and the Pfaffian $\mathcal{M}$   (yellow curve) for a closed system as a function of the Zeeman field $B$ at $\mu$ marked by the white dashed line in (a).   (c) Same as (b) but for an open chain with  $L=300a$ and the Majorana polarization $\mathcal{P}$ indicated by red curve.   Note the absence (presence) of MZMs in panel (b) (panel (c)). Parameters: $\Delta/t =\alpha/t =0.1$ and $\mu/t = 0.1$ in (b,c) indicated by dashed white line in (a).}
	\label{Fig2}
\end{figure*}

Then, given that $P_{\rm L}$ and $P_{\rm R}$ develop opposite signs in the presence of MZMs as anticipated above, we define a new quantity involving their multiplication  as
\begin{equation}
\label{eq:MajoPolTot}
	\mathcal{P} = P_{\rm L}^*\cdot P_{\rm R}= P_{\rm L}\cdot P_{\rm R}^*\,,
\end{equation}
which we refer to as \emph{nonlocal} Majorana polarization because it simultaneously takes into account the behaviour of  $P_{\rm L/R}$. Note that, while $P_{\rm L/R}$ is in general complex, $\mathcal{P}$ is real. Since $P_{\rm L/R}$ are normalized, the maximum absolute value they can take is unity, thus placing the nonlocal Majorana polarization within the range $\mathcal{P}\in[-1,1]$: A value of $\mathcal{P}=-1$ implies that the MZMs are true zero-modes and do not overlap,  whereas a slight deviation from $\mathcal{P}=-1$ corresponds that there is cross-talk between the MZMs or that there is a finite energy splitting of the MZMs energy. Any other value of $\mathcal{P}$ signals a non Majorana state, namely, a topologically trivial state.   With all these characteristics, it is expected that $\mathcal{P}$  not only senses the topological transition but also measures the correlation between MZMs emerging at the edges of the system and, therefore, able to detect the Majorana spatial nonlocality. Following a similar spirit, the nonlocal Majorana polarization given by Eq.\,(\ref{eq:MajoPolTot}) can be easily generalized to higher dimensions, in the same way as   the local Majorana polarization~\cite{Sedlmayr2016Majorana,sedlmayr.bena2015}.
Before going further, it is worth noting that, while the local Majorana polarization~\eqref{eq:MP} is similar to the definition {adopted in Refs.\,\cite{Sedlmayr2016Majorana,sedlmayr.bena2015}}, the  nonlocal polarization introduced  here [Eq.\,\ref{eq:MajoPolTot}]  {goes beyond previous considerations because it accounts for the behaviour on the two halves of the system}  \footnote{It is worth mentioning that the Majorana polarization has been also explored in the many-body context~\cite{Aksenov2020Strong,Tsintzis2022Creating,tsintzis2023roadmap,liu2023enhancing,Souto2024Probing,Samuelson2024Minimal}, with a definition that differs but aims to characterize the same properties of MZMs. Similarly, the Majorana polarization has been used to characterize higher order topological superconductors~\cite{arouca2024mixed}. We note that our definition in Eqs.~\eqref{MPMP} and~\eqref{eq:MajoPolTot} has now been adopted to distinguishing between MBSs and other zero-energy modes in quasicrystals~\cite{Kobialka2024Topological}}. Here we will explore the nonlocal Majorana polarization  $\mathcal{P}$, which will be   obtained for an open chain in real space.  

In what follows, we  investigate the usefulness of the Majorana polarization $\mathcal{P}$ to characterize the topological phase and detect the emergence of MZMs   in the presence of TZESs. For this reason, we  numerically obtain $\mathcal{P}$ and $\mathcal{M}$ in   open and closed finite systems, respectively. Furthermore, we also discuss the energy spectra of the respective systems and contrast with the predictions of  $\mathcal{P}$ and $\mathcal{M}$.  Throughout this work we consider realistic values such that   $\Delta=0.1t$, $\alpha=0.1t$, which correspond to InAs nanowires and Al superconductors \cite{Lutchyn2018Majorana}. For completeness we provide analytic calculation of $\mathcal{P}$ using minimal Kitaev chain in Appendix~\ref{App:MinimalKitaev}. 


\section{Superconducting nanowire}
\label{sec3}
To begin, we analyze the topological properties of a superconducting nanowire modeled by Eq.\,(\ref{Hmodel}) in the absence and presence of scalar disorder, characterized by $V(j)$. In this case, the pair potential ($\Delta$) and chemical potential ($\mu$) are homogeneous all over the nanowire. We explore the Pfaffian and Majorana polarization, which are obtained as discussed in Section \ref{sec2}. Moreover, we also inspect the low-energy spectrum as it reveal the emergence of MZMs. 

\subsection{Superconducting nanowire without disorder}
In the absence of disorder, $V(j)=0$ in Eq.\,(\ref{Hmodel}) and the calculation is carried out numerically. In Fig.\,\ref{Fig2}(a) we present the Pfaffian invariant $\mathcal{M}$ for a closed system as a function of the Zeeman field $B$ and chemical potential  $\mu$, where the blue (red) region correspond to the topological nontrivial (trivial) phase. The white region between the trivial and topological phases correspond the topological phase transition $B=B_{\rm c}$. The green curve in Fig.\,\ref{Fig2}(a) is the Majorana polarization $\mathcal{P}$ obtained for an open system. In Fig.\,\ref{Fig2}(b) and Fig.\,\ref{Fig2}(c) we present the Zeeman dependent low-energy spectrum for a closed and an open system, respectively, which is accompanied by  the Pfaffian $\mathcal{M}$ and Majorana polarization $\mathcal{P}$.

The immediate feature we note in Fig.\,\ref{Fig2}(a) is that the Pfaffian   $\mathcal{M}$ clearly identifies the trivial and topological phases as well as the topological phase transition between them, see red, blue, and white regions. The Pfaffian, however, signals the   topological phase transition for a closed system.
 Notably, by looking at the Majorana polarization $\mathcal{P}$ in the same panel, obtained for open boundary conditions, we observe that it
directly probes the topological phase transition, agreeing perfectly well with the result predicted by the Pfaffian, see green curve between blue and red regions in Fig.\,\ref{Fig2}(a).  This suggests that $\mathcal{P}$ can be also seen as a topological indicator.
By inspecting the low-energy spectrum as a function of the Zeeman field at fixed chemical potential (white dashed line in panel (a)), we find that the Pfaffian obtained for a closed system indeed senses the topological phase transition, as depicted by the red curve jumping from $\mathcal{M}=1$ to $\mathcal{M}=-1$ at $B=B_{\rm c}$ in Fig.\,\ref{Fig2}(b). Note that no MZMs are present in the closed system. In the case of the low-energy spectrum for the open system in Fig.\,\ref{Fig2}(c), the Majorana polarization vanishes for $B<B_{\rm c}$ but, interestingly, reaches $\mathcal{P}=-1$ at the topological phase transition and remains at such value all over the topological phase 
  $B>B_{\rm c}$ where MZMs appear. Thus, the Majorana polarization not only signals the topological phase transition but it also detects the presence of MZMs,   supporting  its interpretation as an alternative real space topological indicator.   

   \begin{figure}[!t]
\centering
	\includegraphics[width=0.49\textwidth]{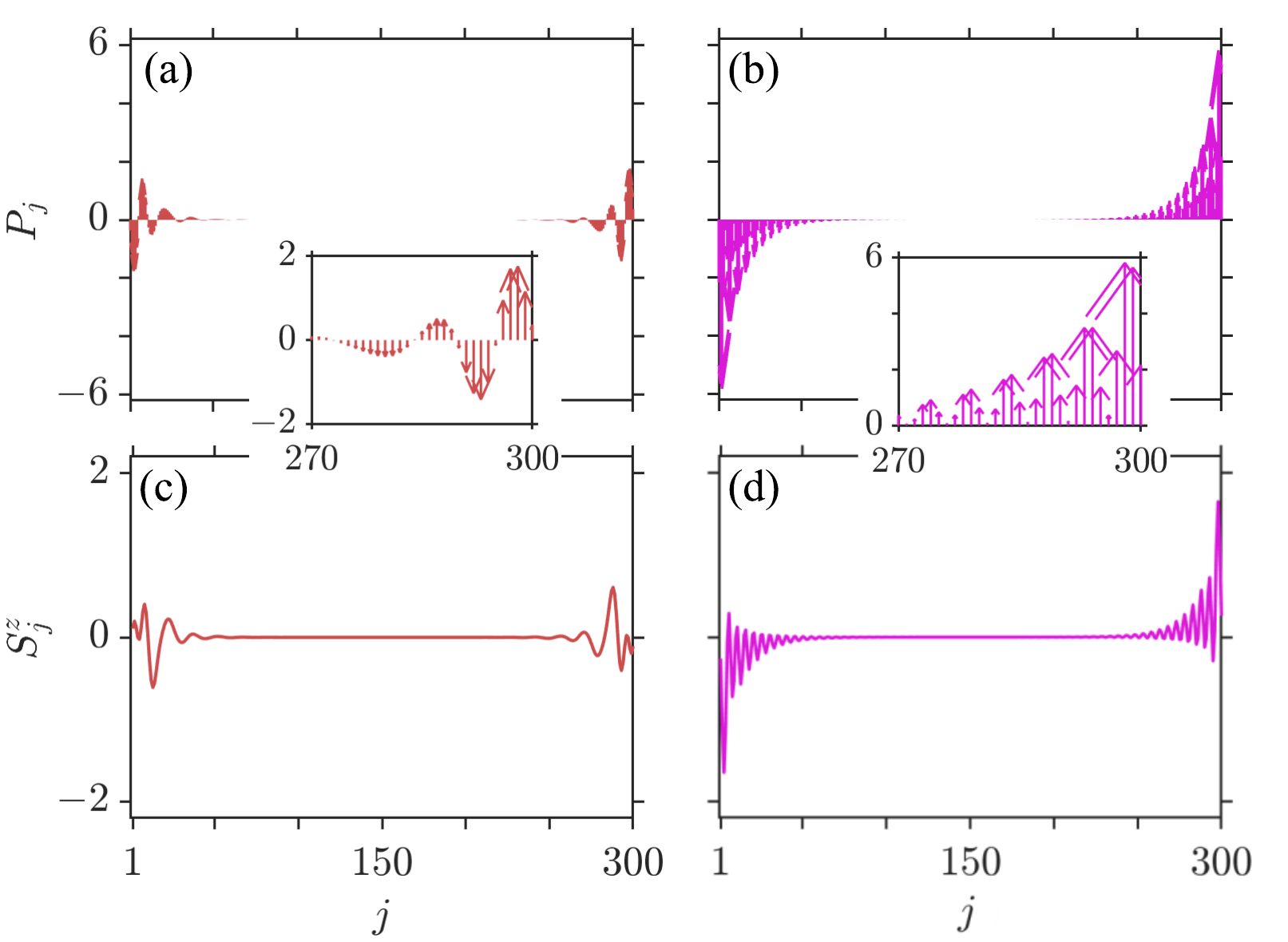}
	\caption{Spatial profiles of the onsite Majorana polarization from Eq.\,(\ref{eq:MP}) (a,b) and the spin projection along $z$ (b,d) in an open chain with $L=300a$.   The colors correspond to the color bars in Fig.~\ref{Fig2}(c). Insets in (a,b) show zoom-in at right edge. In all panels  the vertical axes have been multiplied by $10^2$.}
	\label{Fig3}
\end{figure}

The vanishing (finite) value of the Majorana polarization in the trivial (topological) phase seen in Fig.\,\ref{Fig2}(c) can be understood by inspecting the real space dependence of the onsite Majorana polarization contributions, denoted by $P_{j}$ in Eq.\,(\ref{eq:MP}). In Fig.\,\ref{Fig3} we plot $P_{j}$ as a function of space at Zeeman fields in the trivial and topological phase, indicated by orange and magneta  bars in Fig.\,\ref{Fig2}(c). In the trivial phase $B<B_{\rm c}$, the onsite Majorana polarization $P_{j}$ oscillates in space acquiring  amplitudes that are opposite in sign, which, when summing them up to obtain the left and right Majorana polarizations $P_{\rm L/R}$ in Eq.\,(\ref{MPMP}), give vanishing contribution for the Majorana polarization $\mathcal{P}$ in Eq.\,(\ref{eq:MajoPolTot}). In contrast, in the topological phase $B>B_{\rm c}$, the onsite Majorana polarization $P_{j}$ develops large values at the edges and exhibits an oscillatory decay
 with amplitudes having the same sign up to the middle of the system $L/2$:  $P_{j}<0$ for the left half, while $P_{j}>0$ for the right half. Thus, when summing up  $P_{j}$   to find the left and right Majorana polarizations $P_{\rm L/R}$ in   Eq.\,(\ref{MPMP}), $P_{\rm L/R}$ do not vanish, in contrast to what occurs in the trivial phase. As a result,  the finite value of both $P_{\rm L}$ and  $P_{\rm R}$ prevent the overall vanishing value of $\mathcal{P}=P_{\rm L}P_{\rm R}^*=P_{\rm L}^*P_{\rm R}$ in Eq.\,(\ref{eq:MajoPolTot}), which then explains why $\mathcal{P}$ is finite in the topological phase. When normalizing it to the density-of states, we find $P_{\rm L}=-1$ and $P_{\rm R}=1$, which gives $\mathcal{P}=-1$. Before closing this part, we point out that the  behaviour of the Majorana polarization   follows the corresponding profile of the spin projection, as we indeed observe in Fig.~\ref{Fig3}(c,d) for the  spin density along $z$, obtained as $S^{z}=\Psi^{\dagger}_{j}\sigma_{z}\Psi_{j}$, where $\Psi_{j}$ is the wavefunction associated to the lowest energy state. Of particular interest is that the onsite Majorana polarization follows the same sign of the spin projection, revealing that the negative sign of $\mathcal{P}$ has natural physical interpretation. In sum,  the Majorana polarization is not only a good real space   topological indicator that signals the topological phase transition but it can also be used to identify the formation of MZMs and their spatial overlap.
 
   \begin{figure*}[!t]
\centering
	\includegraphics[width=1\textwidth]{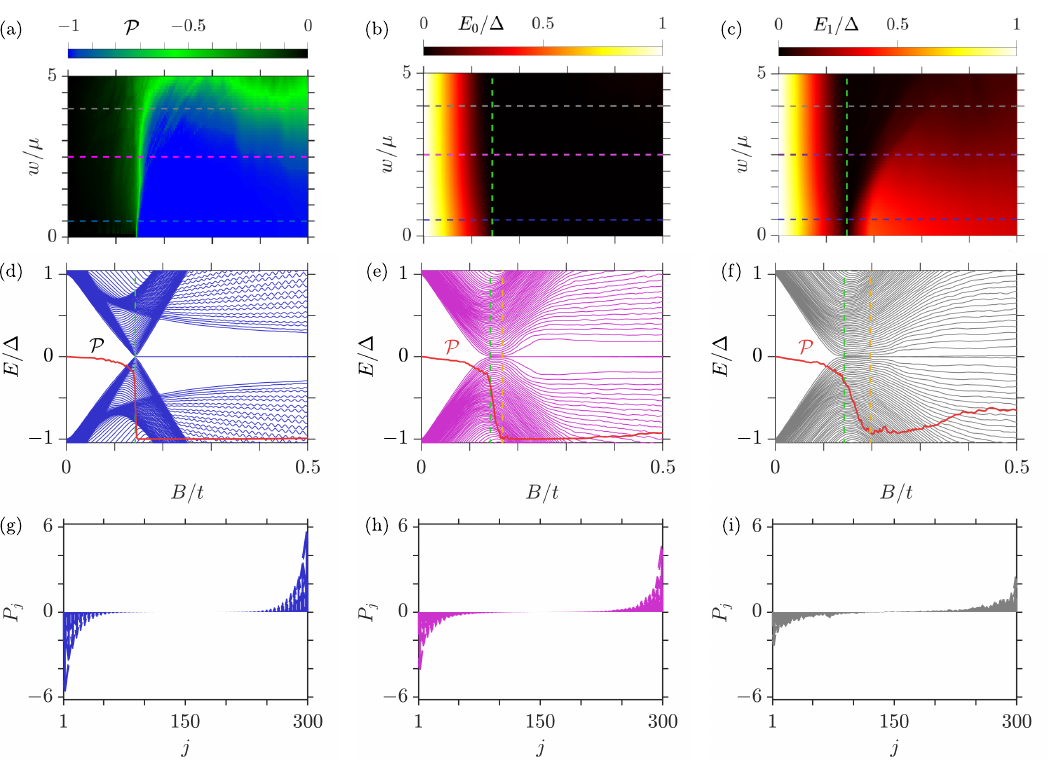}
	\caption{Phase diagram and low-energy spectrum for an open superconducting nanowire under scalar disorder, averaged over 40 disorder configurations. (a) Majorana polarization $\mathcal{P}$ as a function of the Zeeman field and disorder strength $w$ at $\mu/t = 0.1$ (white dashed line in Fig.~\ref{Fig2}(a)). Panels (b,c) show the lowest and first excited energy levels $E_{0,1}$ as a function of $B$ and $w$ at $\mu/t = 0.1$, as in (a). The vertical green dashed line marks the topological phase transition in the absence of disorder.  (d-f) Low-energy spectrum as a function of $B$ at three distinct values of disorder, which correspond to the weak (d) and strong disorder regimes (e,f), marked by  dashed lines in (a-c) with corresponding colors. The Majorana polarization  in (d-f)   is shown by the red curve.	 In (d-f), the  yellow line marks the topological transition point in the   presence of disorder.  (g-i) Onsite Majorana polarization $P_{j}$ in the topological phase at $B= 2B_{c}$ in (d-f) with corresponding colors. The vertical axes of (g-i) have been multiplied by $10^2$. The rest of parameters are the same as the used in Fig.\,\ref{Fig3}. For these parameters strong disorder is obtained when $w/\mu \gtrsim 1.5$.}
	\label{Fig4}
\end{figure*}

\subsection{Superconducting nanowire with disorder}
\label{subsec32}
Having shown the potential of the Majorana polarization to characterize the topological phase and the emergence of MZMs in the previous section, here we explore its robustness under the presence of scalar disorder. The choice of this type of disorder is motivated by the fact that it is very likely to be present in realistic superconductor-semiconductor systems \cite{Zhang2021large,Ahn2021Estimating,Nichele2017Scaling,DasSarma2021Disorder}, see also Refs.\,\cite{aghaee2022inas,sarma2022search}. To model scalar disorder,   we consider random site-dependent fluctuations  characterized by the potential $V(j)$ in Eq.\,(\ref{Hmodel}), 
\begin{equation}
\label{disorderV}
V(j)=\sum_{j\sigma} d_{j\sigma}^{\dagger} [\delta \mu(j)] d_{j\sigma}  + \text{H.c.}\,,
\end{equation}
where $\delta \mu \left(j\right) \in \left[-w,w\right]$ describes site-dependent random fluctuations with  $w $ denoting  the disorder strength, which ensures that
$\langle\delta \mu\rangle=0$. To characterize the amount of disorder, we compare the amplitude of the fluctuations with respect to the chemical potential,  $w/\mu$, since $V(j)$ can be seen as fluctuations around $\mu$. Then, as in the previous section,  we numerically  calculate   the Majorana polarization and the low-energy spectrum including $V(j)$ from Eq.\,(\ref{disorderV}) into the superconducting nanowire Hamiltonian given by Eq.\,(\ref{Hmodel}). 

   \begin{figure*}[!t]
\centering
	\includegraphics[width=1\textwidth]{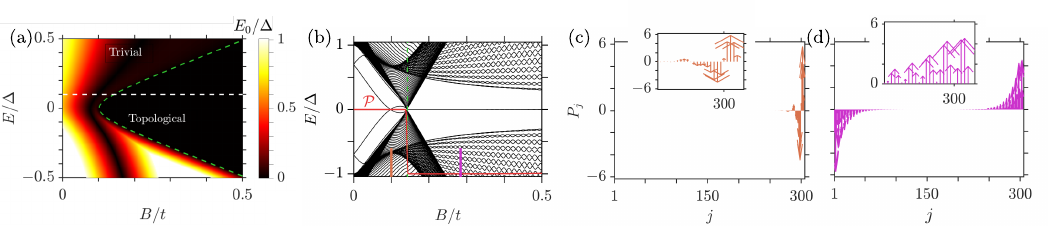}
	\caption{Low-energy spectrum and Majorana polarization in NS junctions. (a) Lowest energy level $E_0$ as a function of the Zeeman field $B$ and chemical potential in the S region $\mu_{\rm S}$ at $\mu_{\rm N}=0.1t$. The dashed green curve corresponds to the topological phase obtained from the Majorana polarization $\mathcal{P}$ for an homogeneous superconductor. (b) Low-energy spectrum and Majorana polarization  $\mathcal{P}$ (red curve) as function of $B$ at $\mu_{\rm S,N}=0.1t$ marked by the dashed white line in (a).  (c,d) Onsite Majorana polarization $P_{j}$ as a function of space at fixed Zeeman fields in the trivial and topological phases, marked by blue and red bars in (b). The vertical axes of (c,d) have been multiplied by $10^2$. Parameters: $L_{\rm N}=4a$, $L_{\rm S}=300a$. Rest of parameters are the same as in Fig.~\ref{Fig4}.}
	\label{Fig5}
\end{figure*}  

In Fig.\,\ref{Fig4}(a-c) we plot the disorder-averaged  of the Majorana polarization $\mathcal{P}$ and the lowest and first excited energy levels $E_{0,1}$ as a function of the Zeeman field $B$ and disorder strength $w$ at fixed chemical potential. In Fig.\,\ref{Fig4}(d-f) we plot the low-energy spectrum and $\mathcal{P}$ as a function of $B$ as three distinct disorder strengths, while in  Fig.\,\ref{Fig4}(g-i) the corresponding onsite $P_{j}$ Majorana polarization as a function of space in the topological phase $B=2B_{\rm c}$. For the chosen parameters, strong disorder is achieved when $w/\mu \gtrsim 1.5$, indicating that (d) is already in the weak-to-intermediate disorder regime, while (e,f) in the strong disorder regime. The first feature to notice is that $\mathcal{P}$ vanishes in the trivial phase for weak-to-intermediate strengths of disorder, develop a sharp transition at the topological phase transition, and 
 acquires a finite value and equal to $-1$ in the topological phase, see black and brightest regions below dashed blue curve in Fig.\,\ref{Fig4}(a). The behavior of $\mathcal{P}$ reflects the formation of MZMs and the gap closing and reopening  in Fig.\,\ref{Fig4}(b,c). At intermediate values of disorder, $\mathcal{P}$ still remains a good indicator, with roughly vanishing values below $B_{\rm c}$ and equal to $-1$ in the topological phase, see  dashed blue curve in Fig.\,\ref{Fig4}(a) and also Fig.\,\ref{Fig4}(d). At strong disorder, $\mathcal{P}$ surprisingly still captures well the topological phase transition even when the visibility of the gap closing and reopening is considerably degraded by an avalanche of TZESs around $B=B_{\rm c}$, see black/dark and bright regions below magenta line in Fig.\,\ref{Fig4}(a-c) and also Fig.\,\ref{Fig4}(d). In fact, $\mathcal{P}\approx -1$ after the topological phase transition (yellow dashed line) in Fig.\,\ref{Fig4}(e), revealing the emergence of well-localized MZMs. This behavior roughly remains at larger values of disorder, with $\mathcal{P}$ being very small below $B_{\rm c}$ but large in the topological phase, see Fig.\,\ref{Fig4}(a-c). It is worth noting, however, that  the Majorana polarization  at very strong disorder in the topological phase ultimately deviates from $-1$ because MZMs acquire a finite spatial overlap that leads to finite energies, see Fig.\,\ref{Fig4}(f).

As in the case without disorder, the finite value of the Majorana polarization $\mathcal{P}$ in the presence of disorder originates from the individual contributions of $P_{j}$ in Eq.\,(\ref{eq:MP}). In fact, in the topological phase, the left and right Majorana polarizations that determine $\mathcal{P}$ develop large values at the edges of the system and decay with an oscillatory fashion towards the bulk of the system, see Fig.\,\ref{Fig4}(g-i). Interestingly, the oscillations of the left (right) Majorana polarization exhibit the same sign, which then lead to a finite  value of $\mathcal{P}$ in the topological phase.  We note that it is the amplitude of $P_{j}$ that gets degraded with the increase of disorder but the spatial profile remains the same. We can therefore conclude that the Majorana polarization $\mathcal{P}$, as defined in Eq.\,(\ref{eq:MajoPolTot}), represents a good topological indicator to characterize the topological phase transition and the emergence of MZMs even in the presence of strong disorder when TZESs appear. 

For completeness in Appendix~\ref{app:MajPol_SingleDis} we present the results for single disorder realization and compare $\mathcal{P}$ and $\mathcal{M}$. The results show the $\mathcal{M}$ fails to capture the topological phase in the presence of TZESs while $\mathcal{P}$ proof to be a robust topological indicator.


\section{SN junctions with trivial zero-energy states}
\label{sec4}
Up to here, we have discussed the Majorana polarization as a tool to characterize the topological phase and MZMs   in single superconducting wires even when disorder-induced TZESs appear. In this part, we consider SN junctions with SOC under a Zeeman field and investigate the potential of the  Majorana polarization to distinguish between MZMs and TZESs. We remark that SN junctions represent another experimentally relevant platform where TZESs emerge due to confinement effects of the N region \cite{PhysRevB.86.100503,PhysRevB.91.024514,PhysRevB.104.134507}. Thus, we take the model given by Eq.~(\ref{Hmodel}) and set $\Delta=0$ in N with chemical potential $\mu_{\rm N}$ and length $L_{\rm N}$, while  S has $\Delta\neq0$ with chemical potential $\mu_{\rm S}$ and length $L_{\rm S}$. Then, we obtain the Majorana polarization and low-energy spectrum as discussed in Section \ref{sec2}; see also  previous Section \ref{sec3}. In Fig.\,\ref{Fig5}(a) we plot the lowest energy level $E_{0}$ as a function of the Zeeman field $B$ and chemical potential in the S region $\mu_{\rm S}$ for a short N region with $\mu_{\rm N}=0.1t$ and relatively long S region. {Therein, the dashed green curve marks the topological phase transition obtained from the Majorana polarization $\mathcal{P}$.} Fig.\,\ref{Fig5}(b) shows the low-energy spectrum and Majorana polarization as a function of $B$ at  $\mu_{\rm N(S)}=0.1t$,  while Fig.\,\ref{Fig5}(c,d) show the onsite Majorana polarization  $P_{j}$ at Zeeman fields that originate  MZMs and TZESs Fig.\,\ref{Fig5}(b).

     \begin{figure*}[!t]
\centering
	\includegraphics[width=1\textwidth]{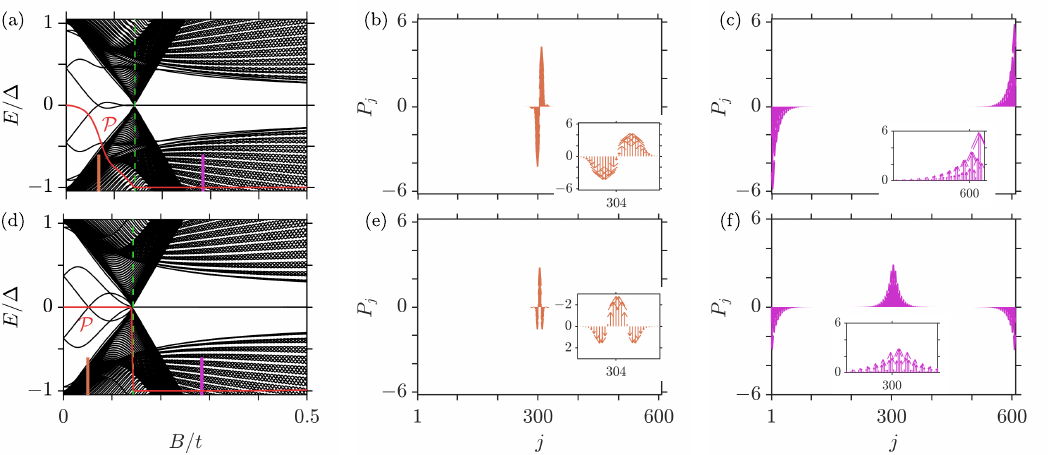}
	\caption{(a,d) Low-energy spectrum and Majorana polarization  in red color as a function of the Zeeman field $B$ for a short SNS Josephson junction at $\phi=0$ and $\phi=\pi$. Panels (b,c)  show the real space dependence of the onsite Majorana polarization $P_{j}$ for $\phi=0$ at Zeeman fields in the trivial and topological phases, marked by blue and red vertical lines in (a). Panels (e,f) show the same  as (b,c) but at $\phi=\pi$. The vertical axes of (b,c,e,f) have been multiplied by $10^2$. {Parameters: $\mu_{\rm N}=0$, $\mu_{\rm S}=0.1t$, $L_{\rm N}=8a$, $L_{\rm S}=300a$, and the rest of parameters considered as in Fig.\,\ref{Fig2}.}}
	\label{Fig6}
\end{figure*}

The first of observation in Fig.\,\ref{Fig5}(a) is that $E_{0}$ reaches zero well before the topological phase transition, which signals the emergence of a TZES whose energy, however, strongly depends on the chemical potential $\mu_{\rm S}$. The formation of the TZES can be also seen in the full  low-energy in Fig.\,\ref{Fig5}(b), where a single level crossing appears below $B_{\rm c}$ (green dashed line) for the chosen parameters. We have verified that this  TZES is located at the interface between the S and N regions, with no counterpart  at the leftmost side of S, as expected for a trivial bound state. This  TZES here appears due to confinement in the finite N region and is thus very sensitive to the choice of parameters  \cite{baldo2023zero,PhysRevB.104.134507}. More TZESs can form when taking longer N regions.  The second observation is that the Majorana polarization, indicated by the red curve in Fig.\,\ref{Fig5}(b), vanishes $\mathcal{P}=0$ all over the trivial phase even at Zeeman fields where  the TZES appears. Remarkably, the Majorana polarization jumps to $\mathcal{P}=-1$  at the topological phase transition (green dashed line) and remains unity all over the topological phase, signalling the emergence of truly MZMs.   The vanishing value of $\mathcal{P}$ in the trivial phase is a result of the onsite Majorana polarization $P_{j}$ acquiring opposite signs in the right half of the wire, which here occurs  even in the presence of TZESs induced by confinement, see Fig.\,\ref{Fig5}(c); {the left half exhibits vanishing $P_{j}$ because there is no bound state on that side; the TZES here localizes at the interface between the right side of S and N}. Remarkably, in the topological phase, the onsite Majorana polarization $P_{j}$ for the left and right halves  decay with an exponentially oscillatory profile that do not change the sign of their amplitudes, resulting in a finite value of $P_{\rm L/R}$ when summing up $P_{j}$ in the left/right halves. Thus, the origin of the finite (vanishing) Majorana polarization in the topological (trivial) phases is the same as the one discussed in the previous two sections.  We can therefore conclude that the Majorana polarization also distinguishes between MZMs and TZESs  in NS junctions.


\section{SNS junctions with four Majorana zero modes}
\label{sec5}
To further demonstrate the utility of the Majorana polarization, here we employ it to characterize the topological phase in short SNS Josephson junctions, where the length of the N region is shorter than the superconducting coherence length. As explained in Section \ref{sec2}, we assume that N has zero pair potential $\Delta=0$,   chemical potential $\mu_{\rm N}$, and length $L_{\rm N}$, while the S regions have pair potential $\Delta\neq0$, chemical potential $\mu_{\rm S}$, and length $L_{\rm N}$. We also consider that there is a finite phase difference between the pair potentials of the superconductors, denoted here by $\phi$. Before we go further, we note that these type of Josephson junctions host two MZMs at $\phi=0$  located at the outer ends of the junction, while four MZMs at $\phi=\pi$, two located at the outer sides and two at the inner sides of the SNS junction, see Fig.\,\ref{Fig1}(c,d) and also Refs.\,\cite{PhysRevLett.108.257001,PhysRevB.96.205425,cayao2018andreev}.  For this reason, we focus on $\phi=0,\pi$ and analyze the Majorana polarization $\mathcal{P}$ for signalling the formation of MZMs.  Since $\phi=0$ and $\phi=\pi$ correspond to a distinct number of MZMs located at different positions, see above, the calculation of $\mathcal{P}$ depends on $\phi=0$. In particular, at $\phi=0$,  the Majorana polarization $\mathcal{P}$ is carried out by dividing the total system into two parts, which serve to calculate the left and right Majorana polarizations $P_{\rm L/R}$ according to Eq.\,(\ref{MPMP}). So, the site index $j$ in Eq.\,(\ref{MPMP}) here runs up to the middle of the N region.  In contrast, at $\phi=\pi$, we obtain $\mathcal{P}$ by taking only half of the total junction, which is then divided into two parts that then are used to find $P_{\rm L/R}$ according to Eq.\,(\ref{MPMP}). We carried out this step  because using the full system at $\phi=\pi$ will lead to double counting and hence vanishing values of Majorana polarization in the presence of the four MZMs. With these annotations, we calculate $\mathcal{P}$ at $\phi=0,\pi$.

In Fig.\,\ref{Fig6}(a,d) we show the low-energy spectrum and Majorana polarization $\mathcal{P}$ as a function of the Zeeman field $B$ at $\phi=0,\pi$ for a short N region with $L_{\rm N}=8a$ and sufficiently long S regions $L_{\rm S}=300a$. Moreover,  Fig.\,\ref{Fig6}(b,c)  and Fig.\,\ref{Fig6}(e,f) depict the spatial dependence of the onsite Majorana polarization $P_{j}$ at $\phi=0$ and $\phi=\pi$, respectively, for  two Zeeman fields in the trivial and topological phases, marked by blue and red bars in (a).  We first note that, for the chosen parameters, the system hosts TZESs below the topological phase transition marked by the   green dashed line, see Fig.\,\ref{Fig6}(a,d). The TZESs at $\phi=0$ have  the same origin as those found in the case of NS junctions and presented in Fig.\,\ref{Fig5}(b). The TZESs at $\phi=\pi$, however, have a distinct origin because  they entirely dependent on the phase difference $\phi$, which also makes them  easily tunable away from zero energy \cite{baldo2023zero}. Nevertheless, these TZESs might also occur in realistic superconductor-semiconductor systems, which is the motivation for the choice of parameters in  Fig.\,\ref{Fig6}. In the topological phase, the low-energy spectrum reveals the emergence of MZMs  at $\phi=0$ and $\phi=\pi$, as expected \cite{PhysRevLett.108.257001,PhysRevB.96.205425,cayao2018andreev}. When inspecting the Majorana polarization $\mathcal{P}$, at $\phi=0$ we observe that it has vanishing values at very small $B$ but smoothly increases up to the topological phase transition but always below $-1$, see red curve in Fig.\,\ref{Fig6}(a); note that $\mathcal{P}$ does not remain constant in the trivial phase even with TZESs as a result of the spatial dependence of $P_{j}$, see below. 
At $\phi=\pi$, the Majorana polarization vanishes all over the trivial phase even in the presence of TZESs originating from $P_{j}$, see red curve in Fig.\,\ref{Fig6}(d). Notably, at the topological phase transition (green dashed line), the Majorana polarization jumps to $\mathcal{P}=-1$ and remains at such value in the entire  topological as $B$ increases, an effect occurring in the presence of two  MZMs at $\phi=0$ and also in the presence of four MZMs at $\phi=\pi$, see red curves in Fig.\,\ref{Fig6}(a,d).  The fact that the Majorana polarization becomes  $\mathcal{P}=-1$ is a strong indicator of the emergence of truly MZMs, which here we show to occur also for MZMs in Josephson junctions.

The finite values of the Majorana polarization $\mathcal{P}$ in the trivial phase at $\phi=0$ can be understood by noting that $P_{\rm L/R}$, which then determine $\mathcal{P}$ as given by Eq.\,(\ref{eq:MajoPolTot}), are obtained  by summing up $P_{j}$ in the left/right half of the total system (the SNS junction). Here, $P_{j}$ has finite values at the junction with opposite signs because of the localization of a TZES. As we see in Fig.\,\ref{Fig6}(b), the left (right) half has  finite $P_{j}$ with the same sign and therefore do not cancel out when summing them up to obtain $P_{\rm L/R}$ and then producing a finite $\mathcal{P}$. This effect does not occur at $\phi=\pi$ in the trivial phase [ Fig.\,\ref{Fig6}(a)] because, being the taken system only the left superconductor and the TZES located at the junction, implies that $P_{j}$ has vanishing values in the left half of the left superconductor. Hence, $P_{\rm L}=0$ and $\mathcal{P}=0$ at $\phi=\pi$ in the trivial phase. In the topological phase, $P_{j}$ reveals intriguing features about MZMs. At $\phi=0$, the outer MZMs exhibit $P_{j}$ with opposite signs   but this situation changes at $\phi=\pi$, where $P_{j}$ for the outer MZMs acquires the same sign. This happens due to the emergence of two extra MZMs at the inner sides of the junction, which exhibit  $P_{j}$ of the same sign. Thus,  the opposite sign profile of $P_{j}$ in each S region is still preserved, with $\mathcal{P}=-1$, in agreement to what we have discussed in previous sections for the single superconducting wire and NS junctions.


\section{Concluding remarks}
\label{sec6}
In conclusion, we have investigated the Majorana polarization for distinguishing between Majorana and trivial zero-energy states in  superconducting systems with spin-orbit coupling and Zeeman field. In particular, we have studied  superconducting nanowires with and without disorder, NS junctions, and SNS junctions. We have shown that the Majorana polarization represents a robust real space topological invariant that is able to detect not only the topological phase transition but also reveals the spatial nonlocality of  Majorana zero modes. Interestingly,  we have demonstrated that the Majorana polarization is robust against strong scalar disorder and serves as an alternative tool to distinguish between Majorana and trivial zero-energy states.

The nonlocal Majorana polarization investigated here also leaves open questions. For instance, the fact that the Majorana polarization is able to identify the topological phase transition and the formation of Majorana zero modes suggests that it captures properties of the emergent topological superconducting state. As such, there seems to be   an intriguing relation between Majorana polarization and emergent superconductivity, which might also require to generalize the definition adopted here to the many-body realm \cite{tsintzis2023roadmap} and therefore deserves further investigation. Another interesting problem is about the measurement of the Majorana polarization.    To access the Majorana polarization, it is necessary to obtain the coherence factors $u$ and $v$, see Eq.~\eqref{eq:MP}. One possible approach to measure such coherence factors is through Andreev   spectroscopy~\cite{Blonder1982Transition}, which can be complemented by  measurements of the local density of states measurements~\cite{Pawlak2016Probing}. Although this task is nontrivial, detecting the Majorana polarization would help characterizing the nature of zero-energy states in Majorana devices.
%
\begin{acknowledgements}
We thank M. Leijnse for important comments on the manuscript.  O. A. A.  acknowledges funding from NanoLund, the Swedish Research Council (Grant Agreement No.~2020-03412) and the European Research Council (ERC) under the European Union's Horizon 2020 research and innovation programme under the Grant Agreement No.~856526. The computations were enabled by resources provided by the National Academic Infrastructure for Supercomputing in Sweden (NAISS) and the Swedish National Infrastructure for Computing (SNIC) at UPPMAX partially funded by the Swedish Research Council (Grant Agreement No.~2020-03412). J. C. acknowledges  financial support from the Swedish Research Council  (Vetenskapsr\aa det Grant No.~2021-04121) and the Carl Trygger’s Foundation (Grant No. 22: 2093).
 \end{acknowledgements}

 
\appendix
 \renewcommand{\thefigure}{A\arabic{figure}}
\setcounter{figure}{0}


\section{Majorana polarization in a two-site Kitaev chain}
\label{App:MinimalKitaev}
In this part we analyze the Majorana polarization in a minimal Kitaev chain consisting of two spin-polarized sites with $p$-wave pairing \cite{Leijnse2012Parity}
\begin{equation}
\label{2KC}
	H_{\rm 2KC}^{\rm}=\sum_{\alpha=L,R}\varepsilon_{\alpha}d_{\alpha}^{\dagger}d_{\alpha}+t d_{\rm L}^{\dagger} d_{\rm R}+\Delta d_{\rm L}^{\dagger} d_{\rm R}^{\dagger}+\text { H.c. }
\end{equation}
where  L and R label the  two (left and right) sites, $\varepsilon_{\alpha}$ is the onsite energy of site $\alpha=L,R$, $t$ is the hopping amplitude, and $\Delta$ the $p$-wave pair potential. The two-site Kitaev chain described by Eq.\,(\ref{2KC}), despite its evident simplicity, holds experimental relevance because it has been recently realized using superconductor-semiconductor systems \cite{dvir2023realization}. Majorana states in Eq.\,(\ref{2KC}) emerge at the so called sweet spot when $\varepsilon_{\rm L,R}=0$ and $\Delta=t$, see Ref.\,\cite{Leijnse2012Parity}. To see the emergence of Majorana states,  we write Eq.\,(\ref{2KC}) in Nambu basis with  $\psi=(d_{\rm L},d_{\rm R},d_{\rm L}^\dagger,d_{\rm R}^\dagger)^T$ and find the eigenvalues and eigenvectors. Then, the lowest energies  in the sweet spot are $E^{0}_{\pm}=0$, and their associated normalized eigenvectors are given by 
\begin{equation}
\label{2KCWave}
\begin{split}
\Psi_{E^{0}_{+}}&=\frac{1}{\sqrt{2}}(0,1,0,1)^{T}\,,\\
\Psi_{E^{0}_{-}}&=\frac{1}{\sqrt{2}}(-1,0,1,0)^{T}\,.
\end{split}
\end{equation}
   \begin{figure}[!t]
	\centering
	\includegraphics[width=0.5\textwidth]{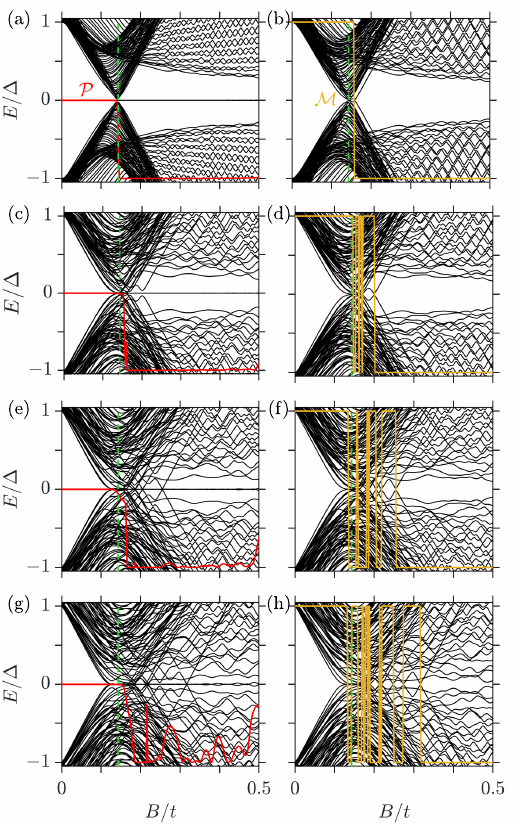}
	\caption{Low energy spectrum, Majorana polarization, $\mathcal{M}$ (left) and Pfaffian, $\mathcal{M}$ (right), for a single disorder realization as functions of Zeeman field, $B$ for several disorder strength values (a,b) $w/\mu = 1$, (c,d) $w/\mu = 2$, (e,f) $w/\mu = 3$, and (g,h) $w/\mu = 4$.}
	\label{FigA}
\end{figure}
By a simple inspection, and reminding the structure of the basis vector $\psi$, we notice that $\Psi_{E^{0}_{+}}$ is located in the right site $\alpha=R$, while  $\Psi_{E^{0}_{-}}$ is located in the left site $\alpha=L$. Hence, the lowest states of the two-site Kitaev chain in the sweet spot exhibit zero energy and are nonlocal. 
Moreover,  $\gamma_{L/R}=\psi.\Psi_{\rm L,R}=\gamma^{\dagger}_{L/R}$ and therefore defines a Majorana operator. These properties allowed to interpret the states defined by Eq.\,(\ref{2KCWave})   as Majorana  quasiparticles; since   there is no topological protection, they are called   poor man's Majorana modes \cite{Leijnse2012Parity}. 

Here we are interested in exploring the expectation value of the particle-hole operator, a quantity identified before as Majorana polarization. In the considered basis, the particle-hole operator is given by $\mathcal{C} = \tau^x$, where $\tau_{i}$ is $i$-th Pauli matrix in Nambu subspace. Then, the expectation value of the particle-hole operator $\mathcal{C} $, or Majorana polarization, for each wavefunction in Eq.\,(\ref{2KCWave}) is given by
\begin{equation}
\begin{split}
p_{\rm L}&=\langle \Psi_{E^{0}_{+}} | \mathcal C| \Psi_{E^{0}_{+}} \rangle=-1\,,\\
p_{\rm R}&=\langle \Psi_{E^{0}_{-}} | \mathcal C| \Psi_{E^{0}_{-}} \rangle=1\,,
\end{split}
\end{equation}
where we have used the subscript L/R because $\Psi_{E^{0}_{+/-}}$ are located in the left/right site. At this point we remark that  the left and right Majorana  state exhibit a Majorana polarization  with  opposite sign. Then, by defining a nonlocal Majorana polarization as  $p=p_{\rm L}^*\cdot p_{\rm R}=p_{\rm L}\cdot p_{\rm R}^*$, it is natural to interpret that   $p=-1$ provides information about the existence of both Majorana-like states perfectly localized in the left and right sites of the two-site Kitaev model given by Eq.\,(\ref{2KC}) in the sweet spot. Although there is no exponential protection, no topological number, the total Majorana polarization  is able to identify the formation of the two MZMs. For obvious reasons, these MZMs have been coined poor man's Majorana modes \cite{Leijnse2012Parity}.

 Away from the sweet spot,   when $\varepsilon_{\rm L/R}=\varepsilon$  and $\Delta \neq t$, the Majorarana polarizations become $p_{\rm L}^{}=-p_{\rm R}^{}={\Delta}/{\sqrt{\Delta^2 +\varepsilon^2}}$, which then leads to $p=p_{\rm L}^{}\cdot p_{\rm R}^{} \approx -1 + 2({\varepsilon}/{\Delta})^2$ for $\varepsilon \ll \Delta$. Hence, as $\varepsilon$  moves away from zero   acquiring finite values, the nonlocal Majorana polarization becomes $p>-1$ signaling that MZMs hybridize. For $\varepsilon\gg\Delta$, the Majorana polarization reads $p\approx -(\Delta/\varepsilon)^{2}$, which practically becomes a very small number, again signalling a strong Majorana hybridization. In the general case, $\varepsilon_{\rm L}\neq \varepsilon_{\rm R}$, MZMs are challenging to realize and the nonlocal Majorana polarization $p$ takes a convoluted form.
\section{Comparison between the Pfaffian and Majorana polarization in disordered systems}\label{app:MajPol_SingleDis}
In the main text we presented disorder-averaged spectrum and showed that the Majorana polarization, $\mathcal{P}$, captures the topological phase accurately, albeit for very strong disorder it deviates from unity due to strong correlation between the MZMs s expected. We also mentioned that the $\mathcal{M}$ fails to distinguish between MZMs and TZESs: due to the fact that $\mathcal{M}$ captures changes in the fermion parity of the ground state. Hence, $\mathcal{M}$ changes for any zero-energy crossing, trivial or nontrivial.

In this Appendix, we compare $\mathcal{P}$ and $\mathcal{M}$ in the presence of plethora TZESs induced by disorder. Fig.~\ref{FigA}(a,c,e,g) shows the low energy spectrum for open chain (black) and $\mathcal{P}$ (red) while Fig.~\ref{FigA}(b,d,f,h) shows the low energy spectrum for closed (black) and $\mathcal{M}$ (gold) for single disorder realization several disorder strength values. At weak disorder (a,b), both $\mathcal{P}$ and $\mathcal{M}$ accurately captures the topological transition. At intermediate, $\mathcal{P}$ (c) still captures the topological phases accurately despite the presence of disorder-induced TZESs while $\mathcal{M}$ (d) changes at every zero-energy crossing. Thus, $\mathcal{P}$ is a good topological indicator in the presence TZESs. The reason for this is that $\mathcal{P}$ only takes into account the MZMs and the correlation between them. Thus, as long as the perturbation does not induce a large cross-talk between the MZMs $\mathcal{P}$ is robust. On the other hand, $\mathcal{M}$ is a bulk quantity as such any changes in fermion parity changes it rendering it a poor topological indicator for inhomogeneous systems. Further increasing the disorder strength $\mathcal{P}$ (e,g) deviates from $-1$ indicating that the topological phase is being destroyed by disorder while the fluctuation in $\mathcal{M}$ increases as the number of zero-energy crossing of trivial states increase with disorder strength (f,h).

Thus, from these results we conclude that $\mathcal{P}$ remains a good topological indicator under the presence of TZESs, thus revealing its potential as a useful tool for the analysis of MZMs in realistic systems.
\bibliography{biblio}
\end{document}